\documentclass[prb,twocolumn,showpacs,preprintnumbers,amsmath,amssymb,superscriptaddress,aps]{revtex4-1}

\usepackage{amsmath,amssymb,amsfont	s} 	
\usepackage{graphicx}
\usepackage{hhline}
\usepackage[latin1]{inputenc}
\usepackage{subfigure}

\usepackage{tabularx}
\usepackage{booktabs}

\usepackage{array}

\newcommand\clearrow{\global\let\rowmac\relax}
\clearrow

\begin{document}

\title{Discovering new two-dimensional topological insulators from computational screening}

\author{Thomas Olsen}
\email{tolsen@fysik.dtu.dk}
\affiliation{Computational Atomic-Scale Materials Design (CAMD), Technical University of Denmark}
\author{Erik Andersen}
\affiliation{Computational Atomic-Scale Materials Design (CAMD), Technical University of Denmark}
\author{Takuya Okugawa}
\affiliation{Computational Atomic-Scale Materials Design (CAMD), Technical University of Denmark}
\author{Daniele Torelli}
\affiliation{Computational Atomic-Scale Materials Design (CAMD), Technical University of Denmark}
\author{Thorsten Deilmann}
\affiliation{Computational Atomic-Scale Materials Design (CAMD), Technical University of Denmark}	
\author{Kristian S. Thygesen}
\affiliation{Computational Atomic-Scale Materials Design (CAMD), Technical University of Denmark}
\affiliation{Center for Nanostructured Graphene (CNG), Department of Physics, Technical University of Denmark}

\begin{abstract}
We have performed a computational screening of topological two-dimensional (2D) materials from the Computational 2D Materials Database (C2DB) employing density functional theory. A full \textit{ab initio} scheme for calculating hybrid Wannier functions directly from the Kohn-Sham orbitals has been implemented and the method was used to extract $\mathbb{Z}_2$ indices, Chern numbers and Mirror Chern numbers of 3331 2D systems including both experimentally known and hypothetical 2D materials. We have found a total of 48 quantum spin Hall insulators, 7 quantum anomalous Hall insulators and 21 crystalline topological insulators. Roughly 75 {\%} are predicted to be dynamically stable and one third was known prior to the screening. The most interesting of the novel topological insulators are investigated in more detail. We show that the calculated topological indices of the quantum anomalous Hall insulators are highly sensitive to the approximation used for the exchange-correlation functional and reliable predictions of the topological properties of these materials thus require methods beyond density functional theory. We also performed $GW$ calculations, which yield a gap of 0.65 eV for the quantum spin Hall insulator PdSe$_2$ in the MoS$_2$ crystal structure. This is significantly higher than any known 2D topological insulator and three times larger than the Kohn-Sham gap. 
\end{abstract}
\maketitle

\section{Introduction}
The concept of topological band theory was initially developed in order to explain the quantum Hall effect, which was observed experimentally in 1980.\cite{Klitzing} The measurements were soon interpreted as a topological effect arising from the phases of Bloch states winding around the boundary of the magnetic Brillouin zone\cite{Thouless1982} and is thus closely related to the $k$-space Berry phase.\cite{Berry1984a} In 1988, Haldane proposed a model system that exhibited the quantum Hall effect without an external magnetic field, but with intrinsically broken time-reversal symmetry.\cite{Haldane1988} Such materials are referred to as quantum anomalous Hall insulators (QAHI) and the first experimental demonstration of the effect was reported in 2013 - 25 years after it was proposed. In the mean time, Kane and Mele had showed that any time-reversal invariant 2D insulator can be characterized by a $\mathbb{Z}_2$ topological index $\nu$. Time-reversal invariant materials with non-trivial topology ($\nu=1$) are known as quantum spin Hall insulators (QSHI) and the effect was observed immediately after a theoretical prediction of the effect in HgTe quantum wells.\cite{Mosfets2013, Konig2007} Subsequently, the concepts have been generalized to bulk 3D systems\cite{Qi2008, Teo2008} and several well-known materials have been shown to comprise examples of topological insulators.\cite{Hasan2010} Most notably, Sb$_2$Te$_3$, Bi$_2$Se$_3$ and Bi$_2$Te$_3$,\cite{Hsieh2009, Zhang2009} but also several 2D materials have been shown to exhibit a non-trivial band topology. In fact, graphene comprised the first theoretical prediction for a QSHI and while it is still believed that graphene has a non-trivial band topology it is practically impossible to verify experimentally due to the small band gap. However, several other 2D materials have been shown to comprise examples of QSHIs. For example, the graphene-like materials silicene,\cite{Ezawa2012, Ezawa2013} germanene,\cite{Zhang2016a, Amlaki2016} and stanene\cite{Tang2014} are all predicted to be QSHIs\cite{Molle2017} and so are several of the transition metal dichalcogenides in the 1T' phase.\cite{Qian2014, Olsen2016a}

While the quantum spin Hall effect has been observed in a wide range of both 2D and 3D materials, the quantum anomalous Hall effect has proven more elusive and has so far only been observed in a few $\mathbb{Z}_2$ topological insulators, where time-reversal symmetry is broken by introducing magnetic impurities.\cite{Chang2013a, Mogi2017a} There has been a few proposals for pristine 2D materials that are predicted to be intrinsic QAHIs by first principles calculations,\cite{Tse2011, Garrity2014, Sheng2017, Wu2017, Chen2017a} but the effect has not yet been confirmed experimentally for any of the materials and the topological properties seem to be somewhat sensitive to the details of the calculations. Moreover, first principles calculations typically only pertain to the case of zero Kelvin, but for 2D materials magnetic order is highly fragile to finite temperature effects and can only be stabilized in the presence of magnetic anisotropy.\cite{Mermin, Torelli2018} Realistic theoretical predictions of novel 2D QAHIs thus have to take into account that the magnetic order must persist at experimentally relevant temperatures - preferably room temperature.

Since the discovery of topological classifications of solids, the field has witnessed a tremendous development of the theoretical concepts, which have been extended to include topological semi-metals,\cite{Armitage2018} topological crystalline insulators\cite{Fu2011, Hsieh2012} and higher order topological effects.\cite{Benalcazar2017a,Benalcazar2017b,Schindler2018a,Schindler2018} In addition to QAHIs and QSHIs we will focus on a particular class of topological insulators in the present work, namely, the topological crystalline insulators where the topology is protected by mirror symmetry in a plane parallel to the 2D material.\cite{Liu2015c, Niu2015} It is particularly easy to understand the topology in this case since the topological index is simply defined as the difference between the quantum anomalous Hall conductance of the two eigenspaces of the Mirror operator. Moreover, the fact that the mirror plane coincides with the plane of the materials implies that any edge that conserves the mirror symmetry will host gapless states that are protected from back-scattering by the topology.

Although topological properties are in principle derivable from the ground state of bulk materials, the observable consequences are very limited in bulk. It is, for example, not possible to distinguish a non-trivial topological state from a trivial one by looking at the band structure alone. However, any interface between materials belonging to different topological classes is guaranteed to host gapless states that are localized at the boundary. Since the boundary states are protected by topology, they are extremely stable and has been proposed as candidates for novel dissipationless electronics circuits. Moreover, since the spin of boundary states are locked to the direction of propagation in QSHIs, such materials are promising for spintronics applications.\cite{Pesin2012} 

A crucial ingredient for operating topological insulators at room temperature is a sizable band gap. Typically, the band gap in topological insulators is determined by the strength of spin-orbit coupling and common values of the gap are on the order of 0.1-0.2 eV. With such small band gaps it becomes hard to maintain full control of the gapless boundary states and the applicability of the materials becomes questionable. It would thus be highly desirable to find new topological insulators with large band gaps. Large-scale screening studies based on first principles computations have previously been performed for 3D materials\cite{Yang2012b, Zhang2018b} and very recently for 2D materials based on experimentally known van der Waals bonded crystals.\cite{Mounet2018} The latter study predicted a novel 2D material in the Jacutingaite prototype to be a QSHI with a band gap of 0.5 eV.\cite{Marrazzo2017a} Remarkably, the PBE gap of the material was only 0.15 eV, but a quasi-particle gap 0.5 eV was obtained with the $G_0W_0$ approximation. It is thus far from obvious that the simple Kohn-Sham gap provides a good estimate of the quasi-particle gap in 2D topological insulators.

Finally, we emphasize that it has not yet been possible to demonstrate the quantum anomalous Hall effect in a pristine 2D material. This is perhaps not so surprising since a magnetically ordered ground state is a minimal requirement for the effect and until very recently magnetic order had not been observed in 2D.\cite{Huang2017a} The discovery of the first pristine 2D QAHI, which exhibits magnetic order at reasonable temperatures thus comprises a tremendous challenge and it is highly likely that theoretical predictions may aid this quest by significantly decreasing the number of relevant materials to investigate.

In the present work we have screened more than 3000 hypothetical 2D materials using first principles simulations and identified novel as well as well-known 2D topological insulators. We have focused on three topological classes: 1) QAHIs, which requires a magnetically ordered ground state and are characterized by the Chern number $C$ that may take any integer value. 2) QSHIs, which require time-reversal symmetry and are classified by the binary $\mathbb{Z}_2$ index. 3) Mirror crystalline topological insulators, where the topology is protected by mirror symmetry and the ground state is classified according to the mirror Chern number $C_M$, which may take any integer value. In order to extract the topological properties we have implemented a full calculation of $k$-space Berry phases that allow us to extract the topological indices in a semi-automated way and does not depend on a mapping to tight binding models through Wannier functions.\cite{Gresch2017}

The paper is organized as follows. In Sec. \ref{sec:hwfs} , we describe the Berry-phase implementation and exemplify how the topological indices are are extracted for the three cases described above. In Sec. \ref{sec:results}, we present the computational details and provide a comprehensive list of all the 2D materials with non-trivial topology that have been found in the screening. We then analyze the topological properties of a few representative materials in more details and investigate the effect of the approximations for exchange-correlation energy and  $G_0W_0$ calculations. In Sec. \ref{sec:discussion} we provide a discussion of the results.

\section{Berry phases and Hybrid Wannier functions}\label{sec:hwfs}
In this section we will briefly introduce the notion of parallel transport and show how it can be applied to obtain the Berry phase matrix of a closed path in $k$-space. We will closely follow the discussions in Refs. \onlinecite{Marzari1997}, \onlinecite{Taherinejad2014}, and \onlinecite{Gresch2017}

\begin{figure}[tb]
   \includegraphics[width=4.5 cm]{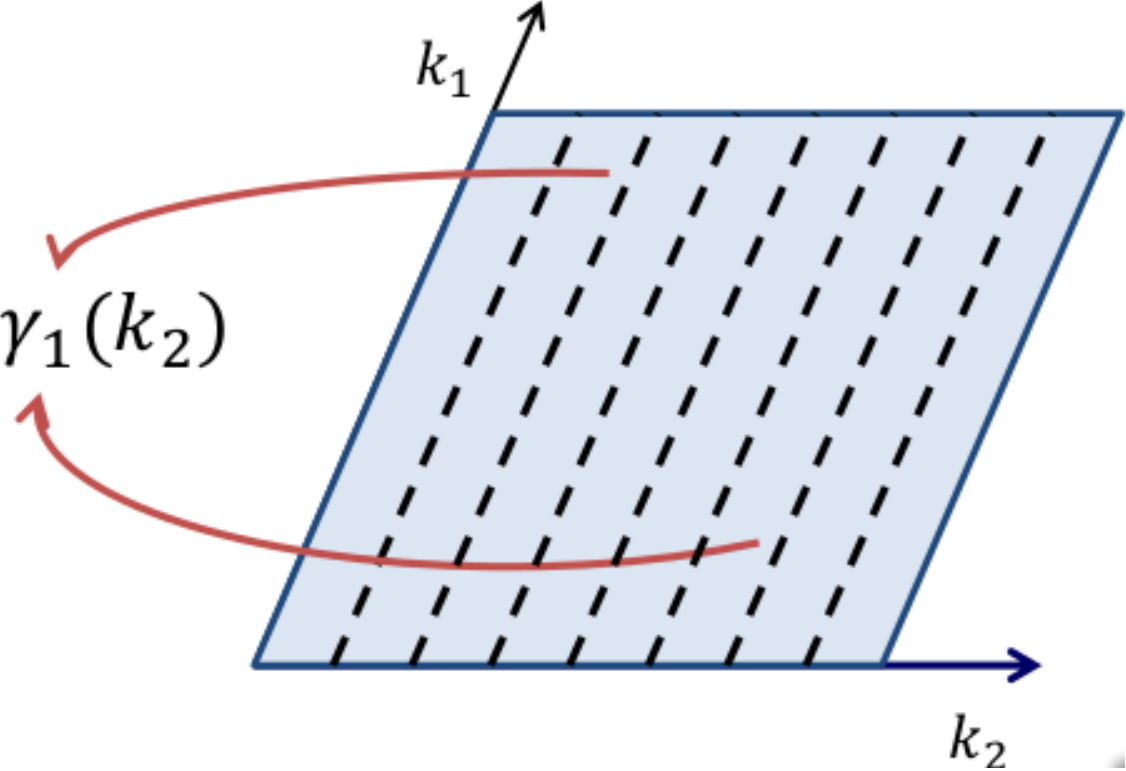}
\caption{The colored area indicate the unit cell in reciprocal space. For each value of $k_2$, the Berry space is calculated by parallel transporting the Bloch states along $k_1$ (indicated by dashed lines).}
\label{fig:bz}
\end{figure}
We consider a minimal unit cell in reciprocal space spanned by the reciprocal lattice vectors $\mathbf{b}_1$ and $\mathbf{b}_2$. A generic point in the reciprocal unit cell can then be written as 
\begin{align}
\mathbf{k}=k_1\mathbf{b}_1+k_2\mathbf{b}_2,
\end{align}
where $0\leq k_i<1$ are the reciprocal fractional coordinates. We wish to calculate the Berry phase obtained by transporting the occupied Bloch states along $k_1$ through the reciprocal unit cell at a fixed value of $k_2$ (See Fig. \ref{fig:bz}). In a numerical treatment of the Bloch Hamiltonian $H(\mathbf{k})$, one obtains a set of occupied eigenstates $|u_{n\mathbf{k}}\rangle$ at different $\mathbf{k}$-points and each set of eigenstates come with an arbitrary set of phases. In fact, any unitary rotation in the space of occupied states leaves the ground state energy invariant and in order to evaluate the phases picked up along a closed path in $k$-space one needs to construct eigenstates with phases that are smooth along the path. This can be accomplished by the so-called parallel transport gauge. For a single occupied state we fix the phase along the path by requiring that $\langle u_\mathbf{k}|\partial_{k_1}u_\mathbf{k}\rangle=0$, which enforces that the change in the state along the path is orthogonal to the state itself. When the Bloch states are calculated at a string of $N$ $\mathbf{k}$-points with fixed $k_2$ and $k_1=0, 1/N, 2/N,\ldots$, the parallel transport condition can be implemented by requiring that $\langle u_\mathbf{k}|u_{\mathbf{k}+\mathbf{b}_1/N}\rangle$ is real at any point along the path. Thus for an initial state at $\mathbf{k}$ on the path we can fix all phases on the path sequentially by the parallel transport condition, which will result in a smooth phase along the path. The phase at $k_1=1$ can then be obtained by imposing the periodic gauge such that $|u_{\mathbf{k}+\mathbf{b}_1}\rangle=e^{-i\mathbf{\hat r}\cdot \mathbf{b}_1}|u_{\mathbf{k}}\rangle$ and the Berry phase is obtained as the phase difference between $k_1=0$ and $k_1=1$.

In the case of multiple occupied bands, the condition is generalized by requiring that the matrix $M_{mn\mathbf{k}}=\langle u_{m\mathbf{k}}|u_{n\mathbf{k}+\mathbf{b}_1/N}\rangle$ is Hermitian. This uniquely fixes the unitary rotation among the occupied states at all points along the path in terms of an initial set of occupied states at particular point $\mathbf{k}$ on the path. This is due to the fact that a single value decomposition yields $M=V\Sigma W^\dag=V\Sigma V^\dag VW^\dag$, where $V$ and $W$ are unitary matrices and $\Sigma$ is diagonal. We can thus take $|u_{n\mathbf{k}+\mathbf{b}_1/N}\rangle\rightarrow WV^\dag|u_{n\mathbf{k}+\mathbf{b}_1/N}\rangle$, which renders $M$ Hermitian and completely fixes the gauge at $\mathbf{k}+\mathbf{b}_1/N$. This procedure is continued along the path until the states at $\mathbf{k}+\mathbf{b}_1$ are obtained using the periodic gauge where $|u_{n\mathbf{k}+\mathbf{b}_1}\rangle=e^{-i\mathbf{\hat r}\cdot \mathbf{b}_1}|u_{n\mathbf{k}}\rangle$. Finally, the eigenvalues of the unitary matrix relating the states at $\mathbf{k}$ and $\mathbf{k}+\mathbf{b}_1$ are the Berry phases acquired by the individual bands.

The method can also be used to obtain the individual states $|\tilde u_{n\mathbf{k}}\rangle$ that are parallel transported without mixing and thus acquire the distinct eigenvalues of the Berry phase matrix.\cite{Marzari1997} Since these are smooth, one may construct hybrid Wannier functions (HWFs) localized along the direction parallel to $\mathbf{a}_1$ as
\begin{align}
|W_{njk_2}\rangle=\int_0^1dk_1e^{-i\mathbf{k}\cdot(\mathbf{\hat r}+\mathbf{\hat R}_j)}|\tilde u_{n\mathbf{k}}\rangle.
\end{align}
Writing $\mathbf{r}=x_1\mathbf{a}_1+x_2\mathbf{a}_2$ with $\mathbf{a}_i\cdot\mathbf{b}_j=2\pi\delta_{ij}$, one can show that\cite{Marzari2012}
\begin{align}\label{eq:wcc}
x_{1,n}(k_2)&\equiv\langle W_{n0k_2}|\hat x_1|W_{n0k_2}\rangle=\gamma_{1,n}(k_2)/2\pi,
\end{align}
with
\begin{align}
\gamma_{1,n}(k_2)=i\int_0^1dk_1\langle \tilde u_{n\mathbf{k}}|\partial_{k_1}\tilde u_{n\mathbf{k}}\rangle.
\end{align}
Except for a factor of $2\pi$, the Berry phases obtained from the parallel transport gauge are thus the charge centers of the HWFs. This construction allows one to calculate various properties associated with the individual phases. For example the spin expectation value
\begin{align}
S_n^{(z)}(k_2)\equiv\langle W_{n0k_2}|\hat S^{(z)}|W_{n0k_2}\rangle,
\end{align}
which will serve as a useful tool for analyzing the topological properties of 2D materials below.

The Berry phases will be smooth functions of $k_2$, which implies that one can track the evolution of the phases while $k_2$ is cycled through the reciprocal space unit cell. The dispersion of the spectrum of Berry phases gives rise to topological classifications as explained below. We will provide examples of \text{ab initio} calculations of the Berry phase spectrum for the three topological classes considered in the present work, but postpone a compilation of the computational details until Sec. \ref{sec:results}.

\subsection{Quantum Anomalous Hall Insulators}
\begin{figure}[tb]
   \begin{center}
   \includegraphics[width=3.0 cm]{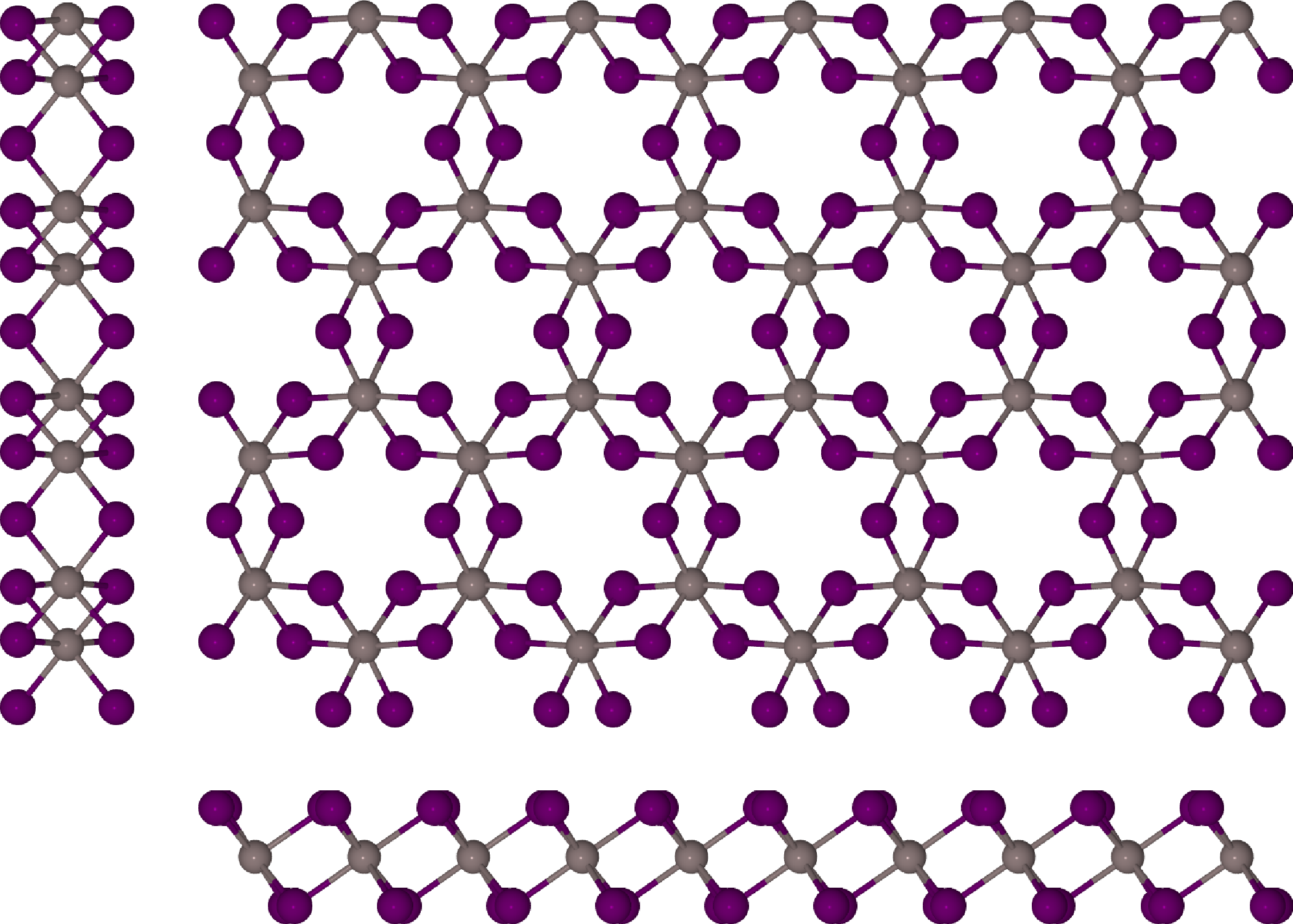}
   \includegraphics[width=2.5 cm]{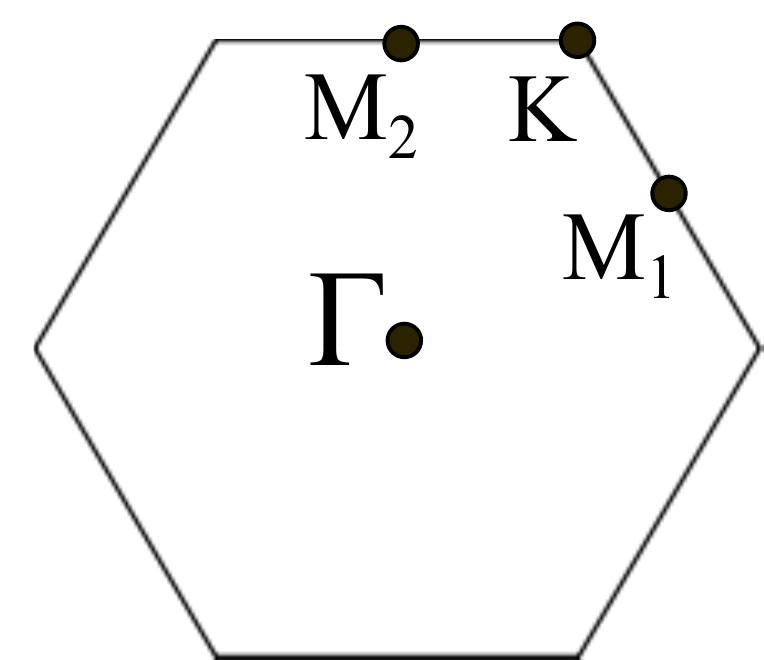}
   \includegraphics[width=8.5 cm]{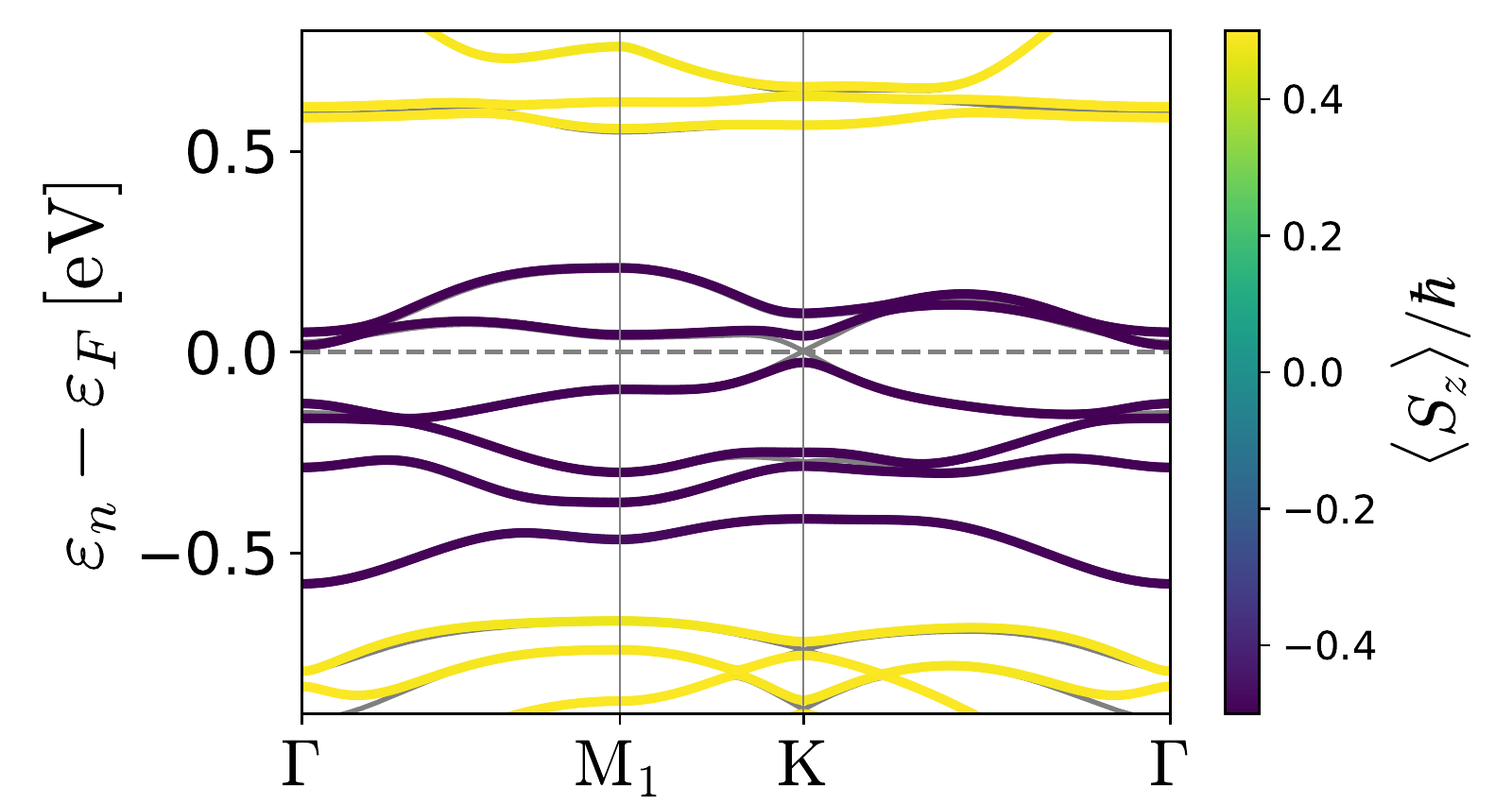}\\
   \includegraphics[width=8.5 cm]{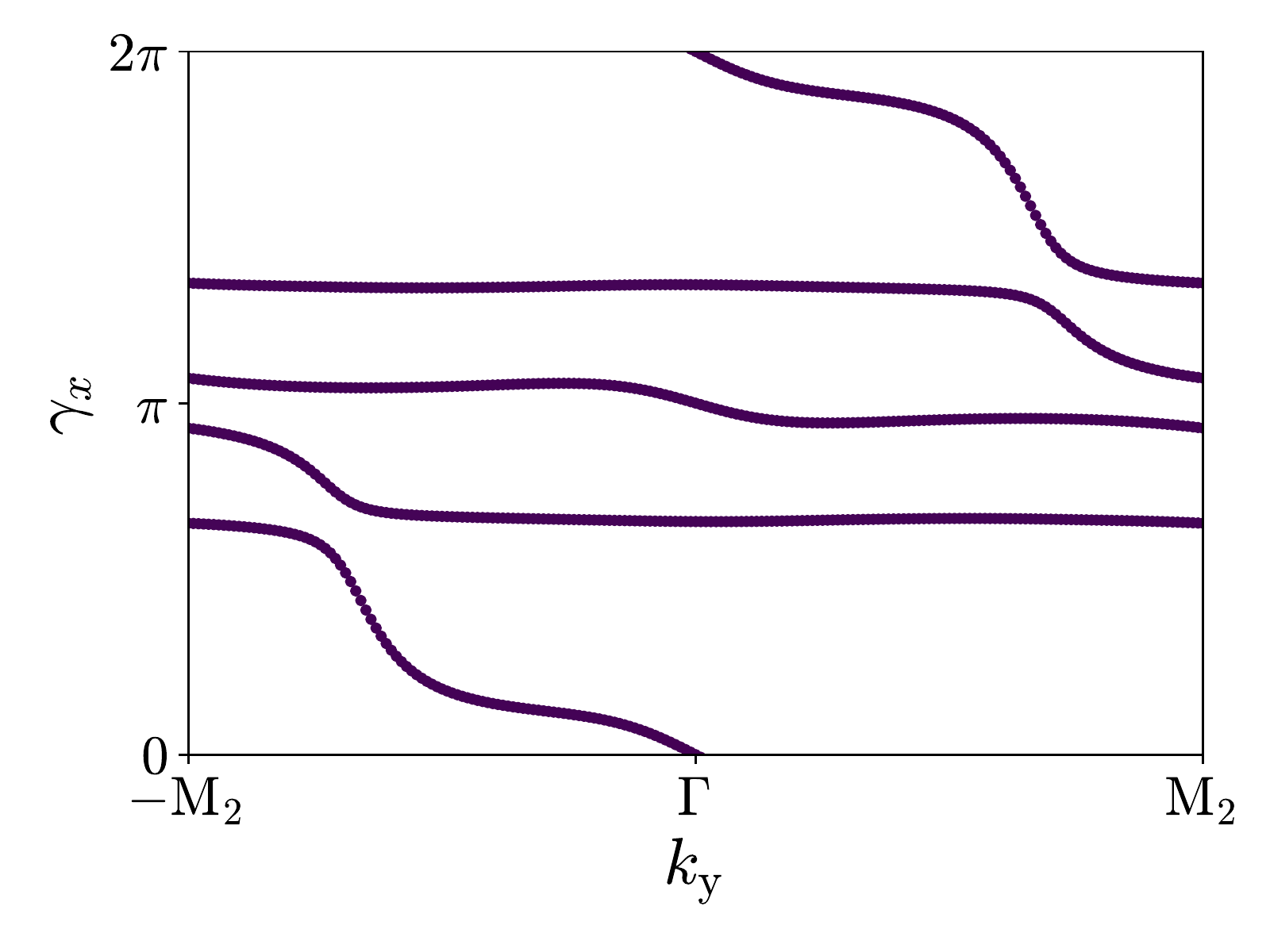}
   \end{center}
\caption{Top: structure and Brillouin zone of FeBr$_3$ in the BiI$_3$ crystal structure. Middle: Band structure of FeBr$_3$ with colors denoting the expectation value of $S_z$. The band PBE gap is 42 meV. Bottom: Berry phases of the 4 highest occupied states of FeBr$_3$ calculated as a function of $k$ in the direction of $\mathrm{M_2}$.}
\label{fig:FeBr3}
\end{figure}
The Hall conductance $\sigma_{xy}$ relates an electric current in the $x$-direction to a uniform field in the $y$-direction by $J_x=\sigma_{xy}E_y$. A finite Hall conductance requires broken time-reversal symmetry and it follows from the Kubo formula that it can be written in terms of the $k$-space Berry curvature $\Omega_{z}(\mathbf{k})$ as\cite{Nagaosa2010}
\begin{align}\label{eq:sigma}
\sigma_{xy}=-\frac{e^2}{h}\int_{BZ}\frac{d^2k}{2\pi}\Omega(\mathbf{k}),
\end{align}
with
\begin{align}\label{eq:Omega}
\Omega(\mathbf{k})=i\sum_{ijn}f_{n\mathbf{k}}\varepsilon_{ij}\partial_{k_i}\langle u_{n\mathbf{k}}|\partial_{k_j}u_{n\mathbf{k}}\rangle.
\end{align}
Here $i,j$ runs over $x$ and $y$, $\varepsilon$ is the two-dimensional Levi-Civita symbol, and $f_{n\mathbf{k}}$ are occupation factors.

For insulators, the integral can be shown to yield an integer known as the Chern number $C$ and the Hall conductance becomes
\begin{align}\label{eq:sigma_C}
\sigma_{xy}=-C\frac{e^2}{h}, \qquad C\in \mathbb{Z}.
\end{align}
In 2D metals, a gap can be opened due to the Landau levels emerging when an external magnetic field is introduced and the value of the Chern number can be controlled by the magnitude of magnetic field. That is the quantum Hall effect. Moreover, as shown by Haldane\cite{Haldane1988} materials with spontaneously broken time-reversal symmetry can exhibit intrinsic quantum Hall effect without an external magnetic field. Such materials are known as quantum anomalous Hall insulators (QAHI) (or Chern insulators) and have a finite Chern number and a non-trivial band topology.

We now briefly discuss how the $k$-space Berry phase calculations outlined above can be related to the Chern number in QAHIs. A constant electric field in the $y$-direction of magnitude $E_0$ can be included in the Bloch Hamiltonian by the substitution $k_y\rightarrow k_y-eE_0t/\hbar$. Clearly the physical properties of the Bloch Hamiltonian are restored after a period $T=2\pi\hbar/eE_0$. If $E_0$ is sufficiently small, the system will evolve adiabatically and the charge transported in the $x$-direction in the time interval is $Q=\int_0^T J_xdt=T\sigma_{xy}E_0=h\sigma_{xy}/e$. Expressing the Hall conductance as $\sigma_{xy}=-Ce^2/h$ we see that the transferred charge is $Q=-eC$. One can thus obtain the Chern number as the number of Wannier charge centers that are transported by a unit cell in the $x$-direction, while $k_y$ is cycled through the reciprocal space unit cell in the backward direction. This argument comprises a modified version\cite{Qi2008} of the explanation originally provided by Laughlin to account for the quantum Hall effect.\cite{Laughlin1981} This also explains the appearance of gapless edge states in QAHIs. For a bulk system an adiabatic cycling of $k_2$ through a reciprocal unit vector will return the system to itself by transferring $C$ charges by one lattice vector. However, in the presence of an edge charges will pile up at the edge and the argument breaks down unless there is a different mechanism that can remove the charges from the edge. One may thus conclude that any edge has to host chiral gapless states that connect the valence band with the conduction band, such that $C$ units of charge are transferred out of the valence bands at the edge while $k_2$ is cycled by a unit.

In Fig. \ref{fig:FeBr3} we show the Berry phases of FeBr$_3$ in the BiI$_3$ crystal structure. Due to Eq. \eqref{eq:wcc}, the vertical axis in the plot can be regarded as the unit cell in the $x$-direction and it is clear that at any horizontal line there will be a total of one state crossing in the downward direction. We can thus calculate the Chern number as the total number of chiral crossings - that is, the number of crossing points with negative slope minus the number of crossing points with positive slope at any horizontal line and we conclude that $C=1$. We have confirmed this result by a direct integration of the Berry curvature $\Omega_z$ over the Brillouin zone.

\subsection{Quantum Spin Hall Insulators}
\begin{figure}[tb]
\begin{center}
   \includegraphics[width=3.0 cm]{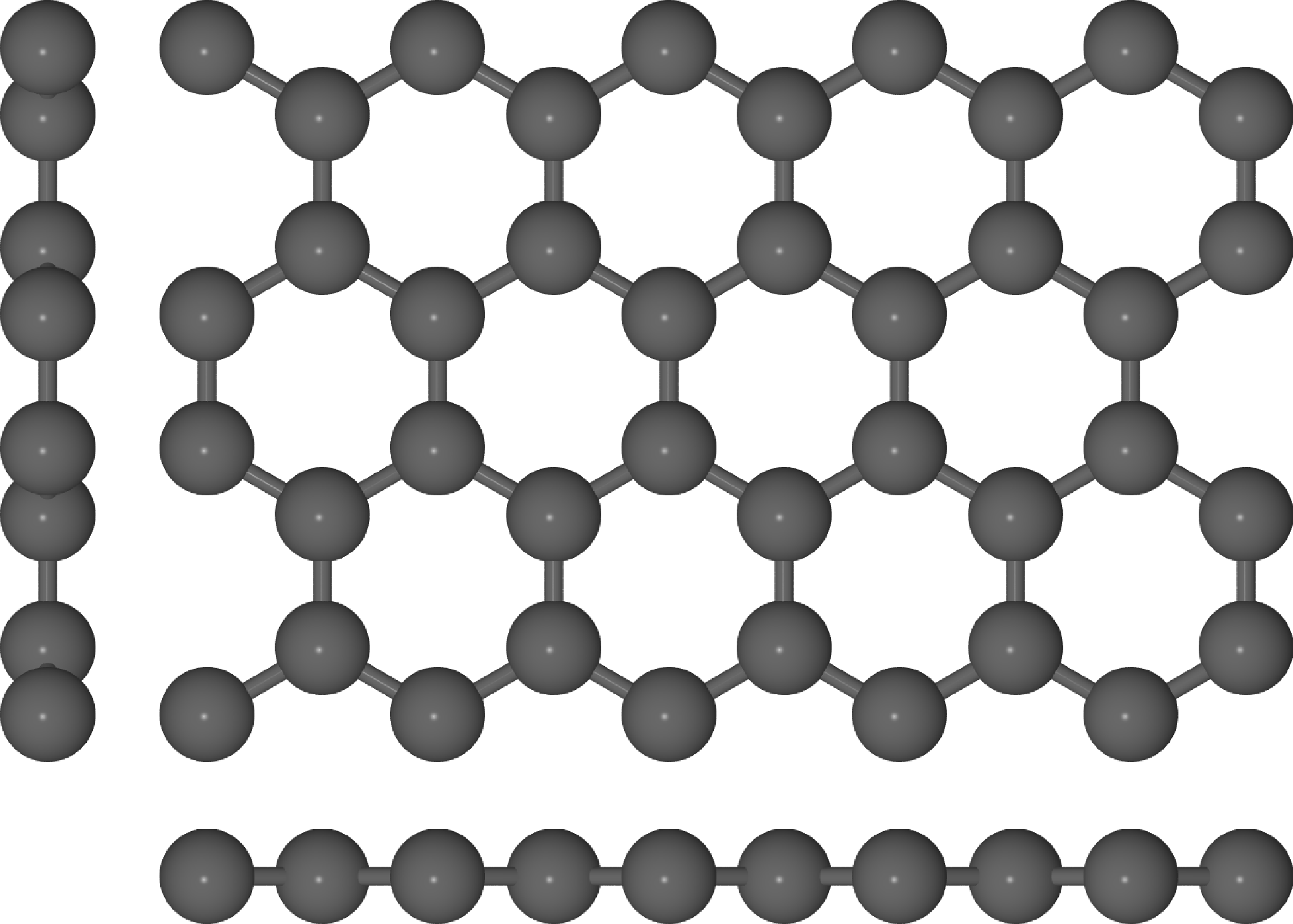}
   \includegraphics[width=2.5 cm]{bz.pdf}
   \includegraphics[width=6.5 cm]{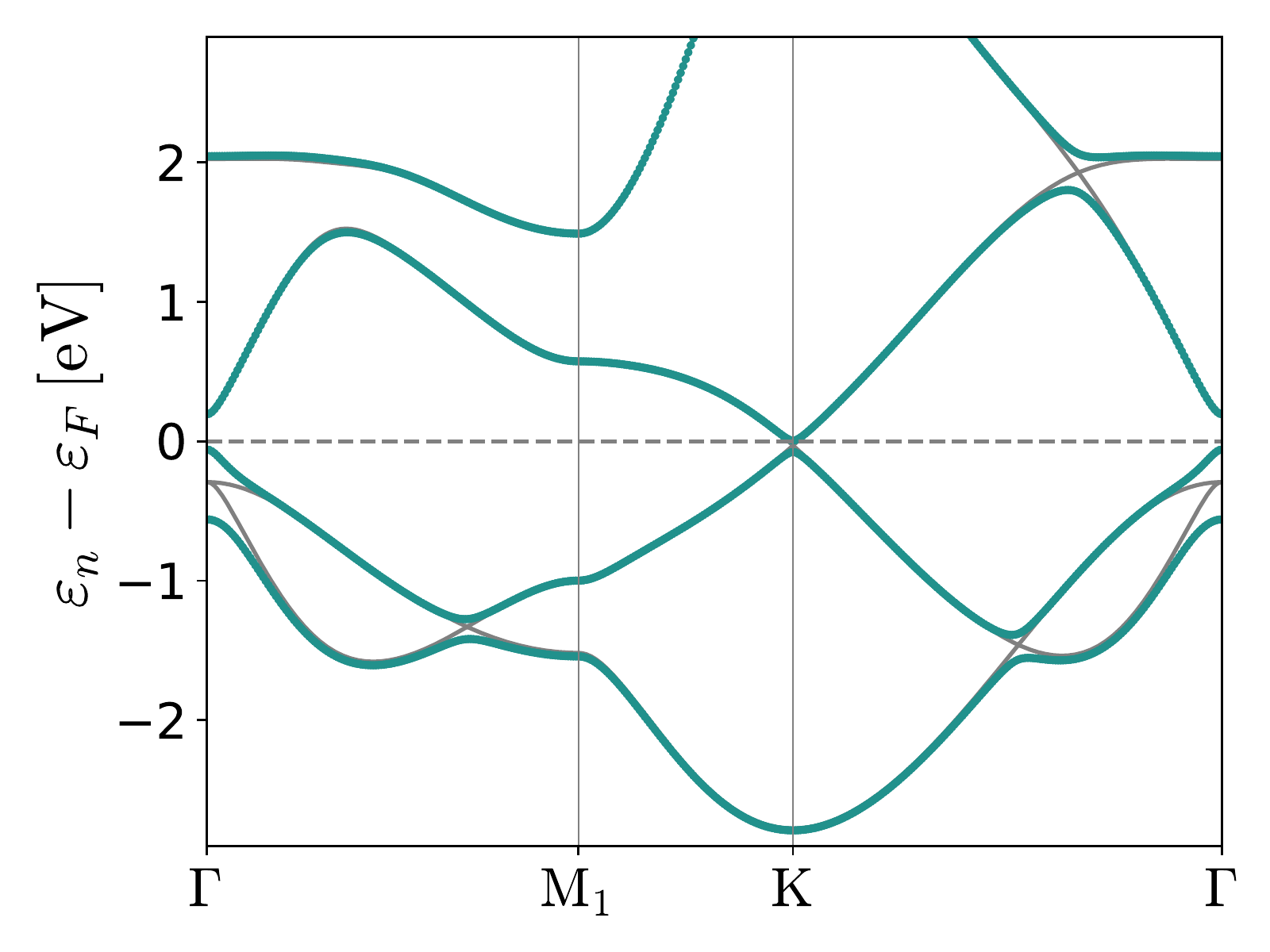}\\
   \includegraphics[width=8.5 cm]{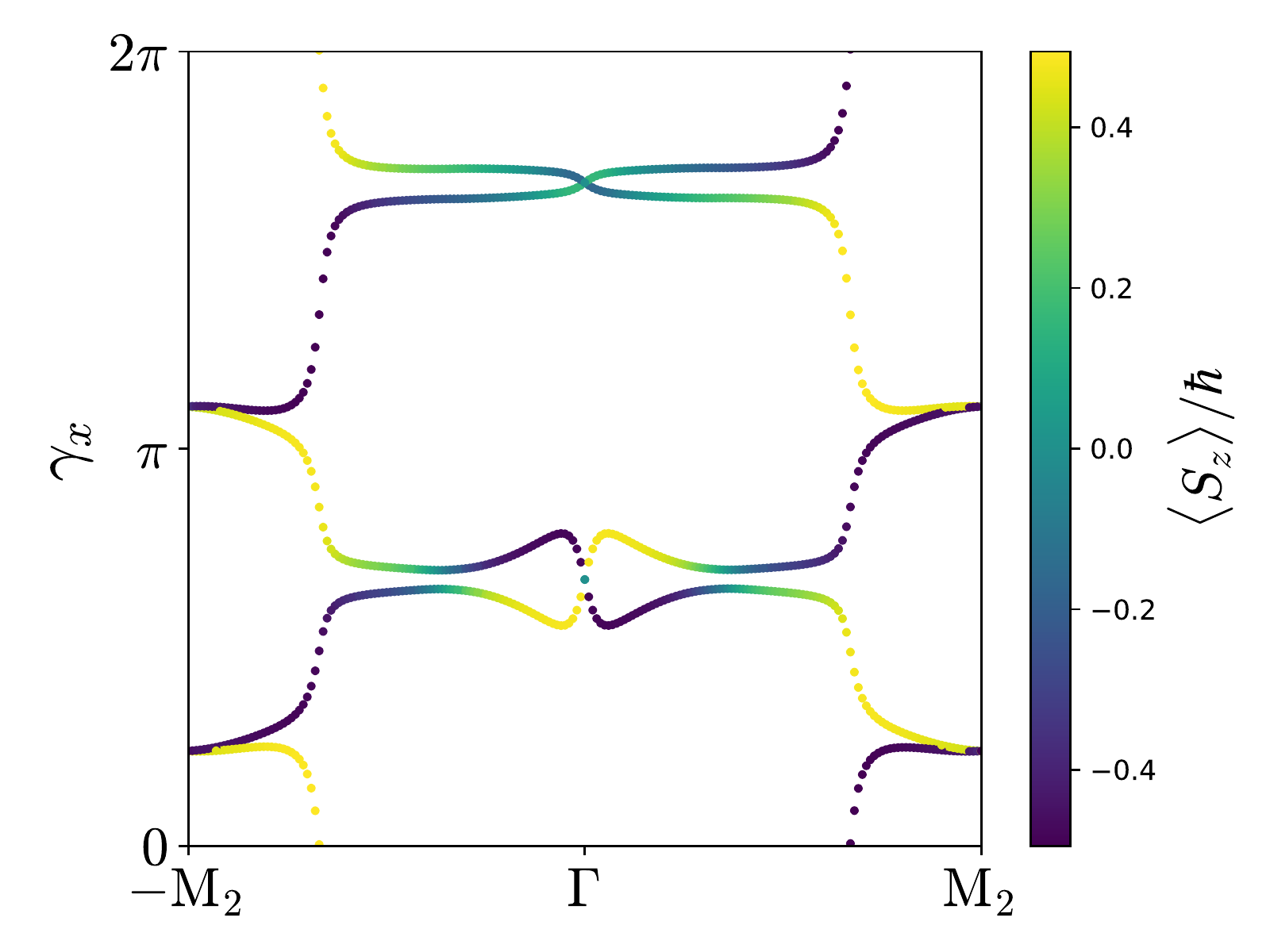}
\end{center}
\caption{Top: structure and Brillouin zone of stanene. Middle: Band structure of Stanene. Bottom: Berry phases of the 4 highest occupied states of stanene calculated as a function of $k$ in the direction of $\mathrm{M_2}$ with colors denoting the expectation value of $S_z$.}
\label{fig:Sn2}
\end{figure}
Materials with time-reversal symmetry must have a vanishing Hall conductance and thus cannot have a non-zero Chern number. From the Berry phase perspective this is due to the fact that any phase at $\mathbf{k}$ will be accompanied by an equal phase at $-\mathbf{k}$, which excludes the possibility of having a finite number of chiral crossings at a horizontal line. Nevertheless, as shown by Kane and Mele\cite{Kane2005,Kane2005a} all 2D time-reversal invariant insulators belong to one of two topological classes and can thus be characterized by a $\mathbb{Z}_2$ topological index. The simplest way to understand the $\mathbb{Z}_2$ index is by considering a system where $S_z$ is a good quantum number. Then one can calculate the Chern numbers of the two spin states and obtain $C_\uparrow=-C_\downarrow$, since $C=C_\uparrow+C_\downarrow=0$. Due to time-reversal symmetry all Berry phases have Kramers degenerate partners at time-reversal invariant points such as $\mathrm{M}$ and $\Gamma$. If one introduces spin-mixing (but time-reversal conserving) perturbations in the Hamiltonian, the two Chern numbers are no longer well-defined but two distinct ways of connecting phases between time-reversal invariant points remains, which gives rise to a $\mathbb{Z}_2$ topological classification. In particular if $C_\uparrow$ is even in the case where $S_z$ is a good quantum number, there will be an even number of Kramers pairs at $\Gamma$ and $M$ that may to hybridize and open a gap in the Berry phase spectrum, once spin-mixing terms are introduced. In contrast, if $C_\uparrow$ is odd, there will be an odd number of Kramers pairs at $\mathrm{M}$ and $\Gamma$ and the Berry phase spectrum must remain gapless when spin-mixing terms are included. In general the $\mathbb{Z}_2$ index distinguishes whether there is an odd or even number of Berry phases crossing any horizontal line in half the Brillouin zone.

In Fig. \ref{fig:Sn2} we show the band structure and Berry phase spectrum for stanene (Sn), which comprises an example of a quantum spin Hall insulator. Due to inversion symmetry in stanene, all bands are doubly degenerate and cannot be colored according to spin. However, the degeneracy of the Berry phases are split due to spin-orbit coupling, which will allow the spectrum to exhibit a single crossing in half the Brillouin zone. Qualitatively, the spin-down electrons are transported upwards, while the spin-up electrons are transported downwards indicating that one would be able to assign spin Chern numbers of $C_\uparrow=-C_\downarrow=1$ if the Hamiltonian of the system could be continuously connected to a Hamiltonian that commutes with $S_z$ without closing the gap.

\subsection{Mirror Crystalline Topological Insulators}
\begin{figure}[tb]
   \begin{center}
   \includegraphics[width=3.0 cm]{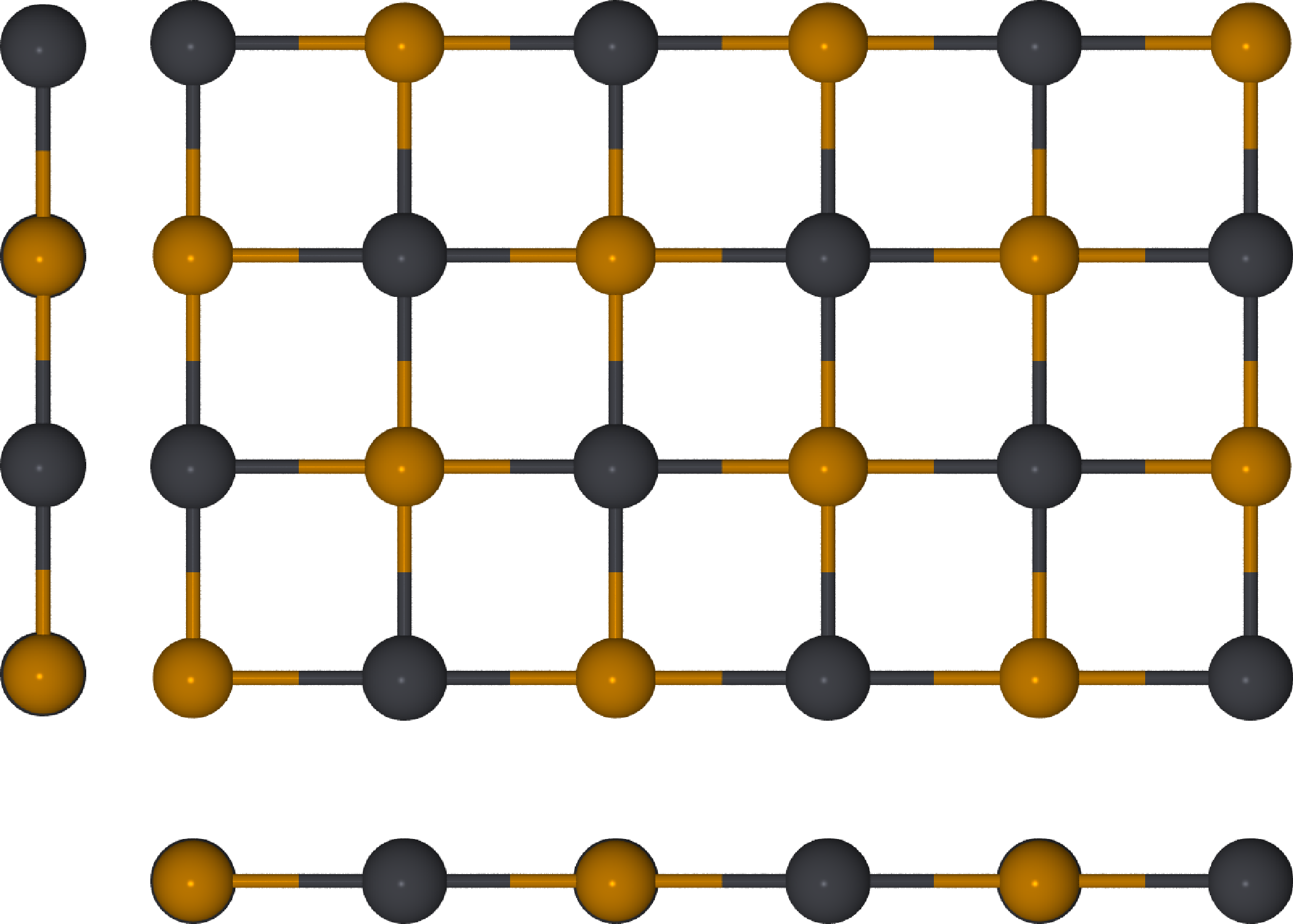}
   \includegraphics[width=2.5 cm]{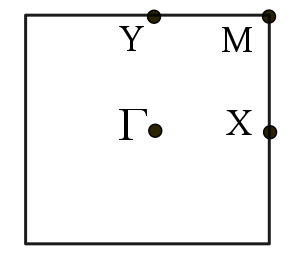}
   \includegraphics[width=6.5 cm]{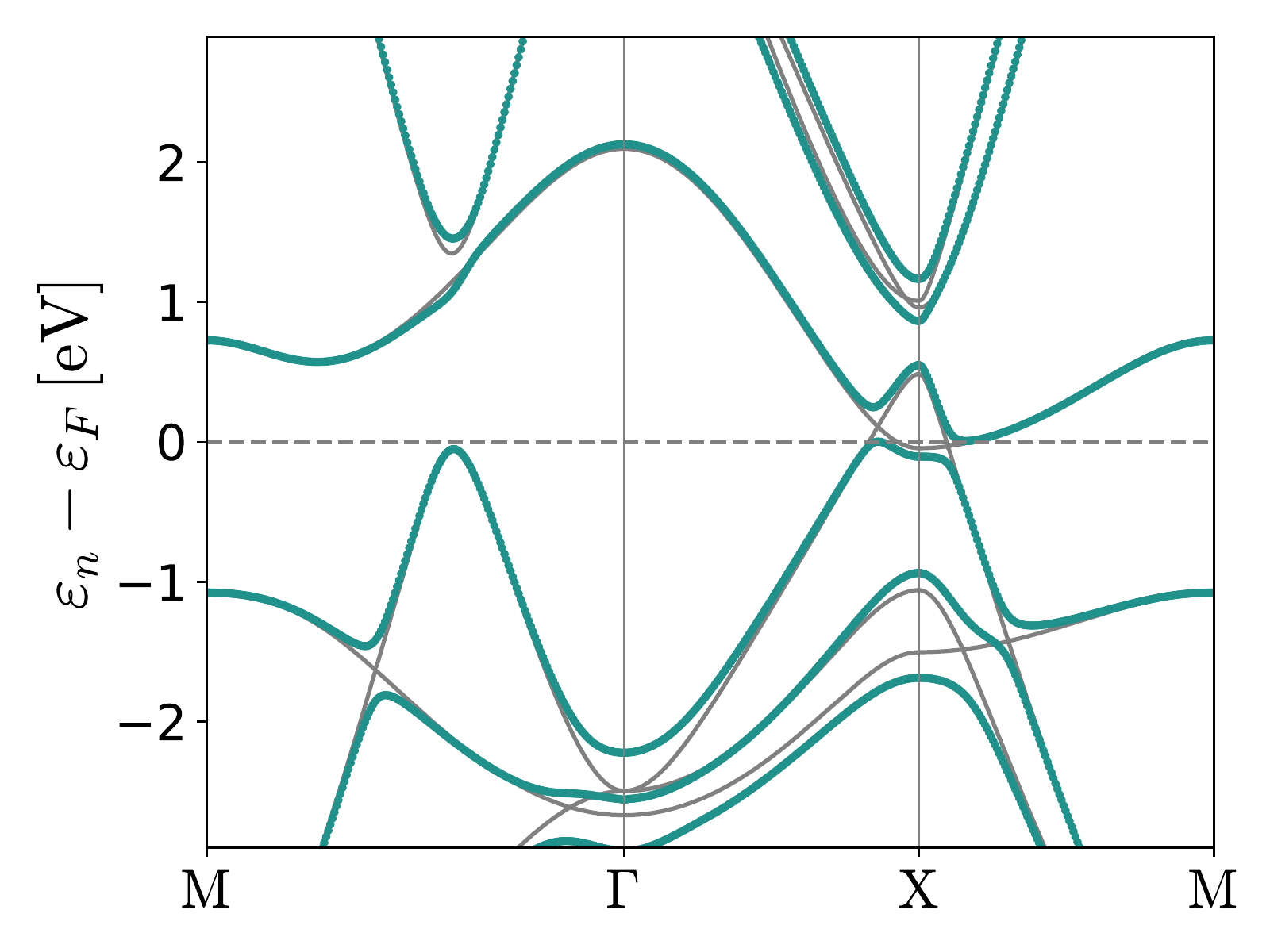}\\
   \includegraphics[width=8.5 cm]{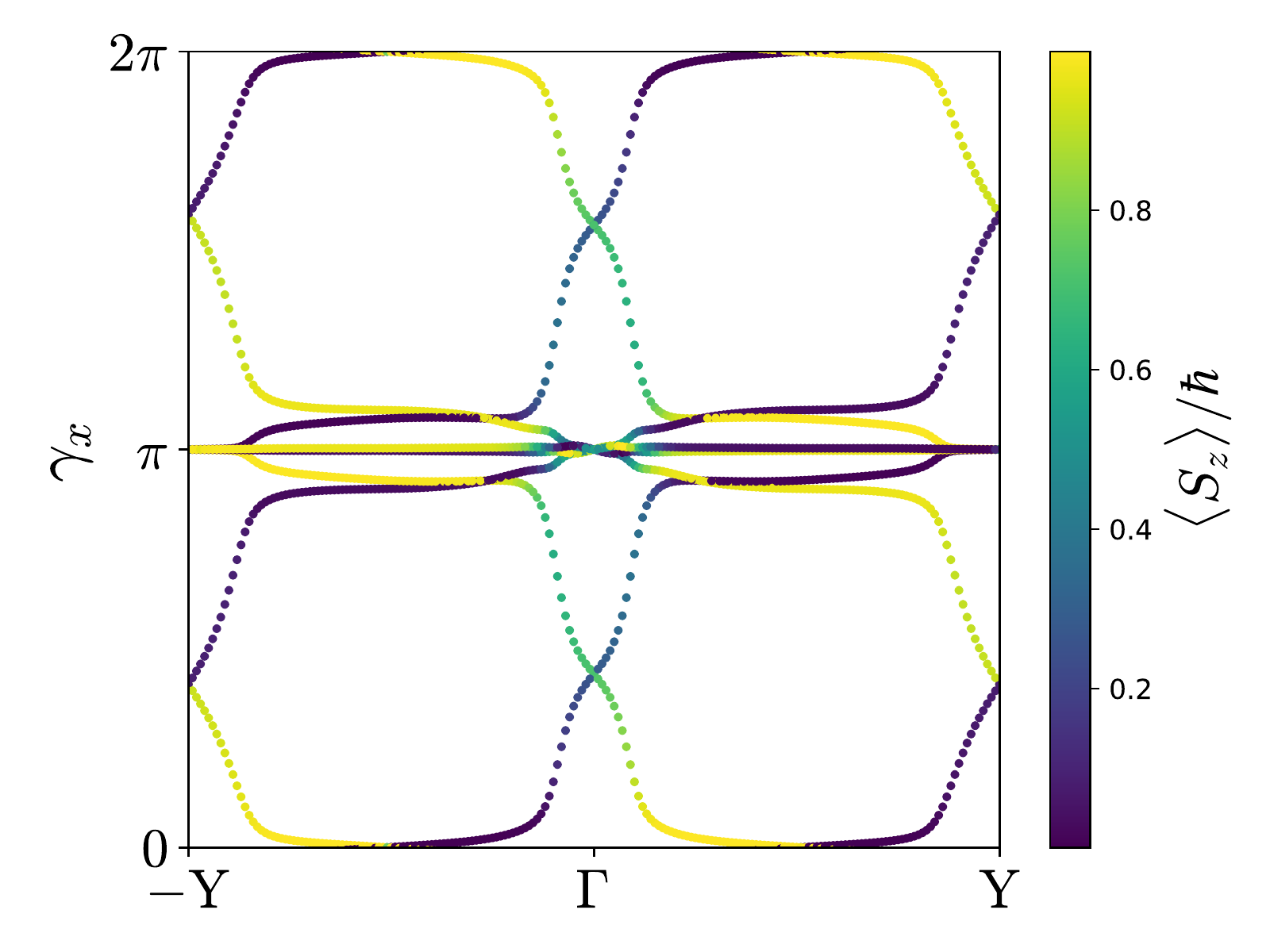}
   \end{center}
\caption{Top: structure and Brillouin zone of SnTe in the PbSe crystal structure. Middle: Band structure of SnTe. Bottom: Berry phases of the 8 highest occupied states of FeBr$_3$ calculated as a function of $k$ in the direction of $\mathrm{Y}$ with colors denoting the expectation value of $S_z$.}
\label{fig:SnTe}
\end{figure}
As shown by Fu,\cite{Fu2011} crystal symmetries alone may give rise to a topologically non-trivial band structure. However, the consequences (gapless boundary states) are only observable at edges or surfaces that conserve the crystal symmetry. In this respect, mirror symmetry comprises a particularly simple type of crystal symmetry that gives rise to an integer topological classification. Whenever a material has mirror symmetry all the occupied states may be labeled according to their mirror eigenvalues $\pm i$ and one can define Chern numbers $C_\pm$ for these subsets of bands. The total Chern number is then $C=C_++C_-$, but we might as well consider the $\mathbb{Z}\times\mathbb{Z}$ classification based on $C_+$ and $C_-$. In the case where the total Chern number vanishes the topology can be specified by the mirror Chern number $C_M=(C_+-C_-)/2$, which is readily verified to be an integer.

For 2D materials, mirror symmetry in the plane of the material plays a special role, since any clean edge of the material will conserve the symmetry. A 2D material with non-vanishing mirror Chern number is thus guaranteed to host gapless edge states and gapless boundary states if the material is interfaced with another material that exhibits mirror symmetry in the plane. Strictly speaking, the interface or edge itself may break the symmetry by reconstruction of adsorption of atoms or molecules and as such the gapless edge states are more fragile than in the case of QSHI where the topology is defined by time-reversal symmetry. If $C=0$ an edge will host $C_M$ edge states of positive chirality and $C_M$ edge states of negative chirality. In Fig. \ref{fig:SnTe} we show the Berry phase spectrum of SnTe, which is an example of a mirror crystalline topological insulator with $C_M=2$.\cite{Liu2015c} We note that spinorial part of the mirror operator in the $z$-plane is represented as $i\sigma_z$ and the mirror eigenvalues thus closely follow the spin of the hybrid Wannier functions. This is evident in Fig. \ref{fig:SnTe}, where the spin up states are largely transported downward ($C_+=2$) and the spin down states ($C_-=-2$) are largely transported upward. This is of course only a qualitative argument, since the mirror operator also affects the orbital part, but the spin structure of the hybrid Wannier functions does allow one to obtain an intuitive picture of the (mirror-resolved) charge transport.

It is interesting to note that in the presence of time-reversal symmetry, the $\mathbb{Z}_2$ index can be obtained as $C_M$ mod 2. This is due to the fact that time-reversal symmetry enforces mirror symmetry on the Berry phase spectrum around the vertical axis at the $\Gamma$-point. In particular, we can represent the time-reversal symmetry operator as $T=\sigma_yK$, where $K$ denotes complex conjugation and the mirror symmetry operator by $M_z=i\sigma_z$. Due to the fact that $\sigma_y$ and $\sigma_z$ anti-commute, it follows that if $|u_+\rangle$ is an eigenstate of $M_z$ with eigenvalue $+i$ then $T|u_+\rangle$ will be an eigenstate of $M_z$ with eigenvalue $-i$. The Berry phase spectrum of the negative eigenvalue sector can thus be obtained from that of the positive mirror eigenvalues by reflection through the vertical line at $\Gamma$. This results in an odd number of crossings in half the Brillouin zone if $C_M$ is odd and an even number of crossings if $C_M$ is even. Materials with odd mirror Chern number and time-reversal symmetry thus have a dual topological character such that the edge states will be remain protected if either mirror symmetry or time-reversal symmetry is broken. This can be regarded as a 2D analogue of the dual topological character in Bi$_2$Se$_3$.\cite{Rauch2014,Olsen2016c} In that case the strong topological index can be calculated from the Chern-Simons axion coupling $\theta$, which is restricted to the values of $0$ and $\pi$ in the presence of either mirror symmetry or time-reversal symmetry.

\section{Results}\label{sec:results}
\subsection{Computational details}
All the calculations presented in the this work were based on the Computational 2D Materials Database (C2DB), which currently contains DFT calculations for 3331 2D materials.\cite{Haastrup2018} The calculations were performed with the electronic structure software package GPAW,\cite{Enkovaara2010a} which is based on the projector augmented wave method\cite{blochl} combined with the Atomic Simulation Environment (ASE).\cite{Larsen2017} All materials have been fully relaxed with the PBE functional\cite{pbe} and treated according to a strict work flow and a wide range of properties are calculated for materials that are both dynamically and thermodynamically stable. We refer to Ref. \onlinecite{Haastrup2018} for details on the calculations. The database can be browsed online at https://c2db.fysik.dtu.dk or the full database can be downloaded from the repository.

As a first screening for 2D materials with non-trivial topology, we have sorted the C2DB for insulators with direct PBE band gaps below 0.7 eV. This primary criterion is based on the fact that the topological gaps are driven by spin-orbit coupling and we do not expect spin-orbit coupling to open band gaps by more than 0.7 eV. For all these materials, we carried out the parallel transport described in Sec. \ref{sec:hwfs} and identified topological insulators as the materials with a gapless Berry phase spectrum. The procedure is highly convenient for automated screening, because no reference to the type of topological insulator is needed for the calculations and one can simply sort out whether a given materials is a QAHI, QSHI or a crystalline topological insulator (protected by mirror symmetry) afterwards.

The only non-standard ingredient in the procedure is the calculations of the matrix elements
\begin{align}
M_{mn}(\mathbf{k},\mathbf{k}+\delta\mathbf{k})&=\langle u_m(\mathbf{k})|u_n(\mathbf{k}+\delta\mathbf{k})\rangle\notag\\ 
&=\langle \psi_m(\mathbf{k})|e^{-i\delta\mathbf{k}\cdot\mathbf{\hat r}}|\psi_n(\mathbf{k}+\delta\mathbf{k})\rangle,
\end{align}
which is needed for the parallel transport algorithm. Within the PAW formalism the all-electron wavefunctions are written as
\begin{align}
|\psi_n\rangle=|\tilde\psi_n\rangle+\sum_{ai}\langle \tilde p_i^a|\tilde\psi_n\rangle\Big[|\phi^{a}_i\rangle-|\tilde\phi^{a}_i\rangle\Big],
\end{align}
where $|\tilde\psi_n\rangle$ are soft pseudo-wavefunctions, $|\phi_i^a\rangle$ are atomic orbital of atom $a$ and $|\tilde p_i^a\rangle$ are projector functions satisfying $\langle\tilde p_i^a|\tilde\phi_j^a\rangle=\delta_{ij}$. Denoting the position of atom $a$ by $\mathbf{r}_a$ the matrix elements can thus be written as
\begin{align}\label{eq:M}
  M_{mn}(\mathbf{k},\mathbf{k}+\delta\mathbf{k})&=\langle\tilde\psi_m(\mathbf{k})|e^{-i\delta\mathbf{k}\cdot\mathbf{\hat r}}|\tilde\psi_n(\mathbf{k}+\delta\mathbf{k})\rangle\notag\\
  &+\sum_{aij}e^{-i\delta\mathbf{k}\cdot\mathbf{r}_a}\langle\tilde\psi_m(\mathbf{k})|\tilde p_i^a\rangle\\
  &\times\Big[\langle\phi_i^a|\phi_j^a\rangle-\langle\tilde\phi_i^a|\tilde\phi_j^a\rangle\Big]\langle\tilde p_j^a|\tilde\psi_n(\mathbf{k}+\delta\mathbf{k})\rangle,\notag
\end{align}
where we assumed that $\langle\phi_i^a|e^{-i\delta\mathbf{k}\cdot\mathbf{\hat r}}|\phi_j^a\rangle=e^{-i\delta\mathbf{k}\cdot\mathbf{r}_a}\langle\phi_i^a|\phi_j^a\rangle$, since the partial waves $|\phi_i^a\rangle$ are localized at the atom $a$ and $e^{-i\delta\mathbf{k}\cdot\mathbf{r}}$ is a slowly varying function when $\delta\mathbf{k}$ is small. All quantities entering in Eq. \eqref{eq:M} are calculated during standard DFT calculations with GPAW and are thus readily available for Berry phase calculations.

\subsection{Overview of topological insulators in C2DB}
In Tabs. \ref{tab:table1} and \ref{tab:table2} we provide an overview of all the topological insulators found in the screening. We emphasize again that all these were simply identified by looking for materials with a gapless Berry phase spectrum. The topological indices $\nu$, $C$, and $C_M$, relevant for QSHIs, QAHIs and MCTIs respectively, were then assigned to the different materials afterwards. 

\begin{table}[tb]
  \begin{center}
    \label{tab:table1}
    \begin{tabular}{l|c|c|c|c|}
      Material & Prototype & Topology & KS gap [meV] \\ 
      \hline
      C$_2$\cite{Kane2005a}    & C$_2$   & $\nu=1$, $C_M=1$ & 0.3 \\ 
      Si$_2$\cite{Ezawa2015}   & C$_2$   & $\nu=1$ & 1.6   \\ 
      Ge$_2$\cite{Ezawa2015}   & C$_2$   & $\nu=1$ & 25     \\ 
      Sn$_2$\cite{Ezawa2015}   & C$_2$   & $\nu=1$ & 65     \\ 
      SnF\cite{Xu2013}      & CH      & $\nu=1$ & 316    \\ 
      HgSe\cite{Li2015}     & GeSe    & $\nu=1$ & 90    \\ 
      HgTe\cite{Li2015}     & GeSe    & $\nu=1$ & 156   \\ 
      MoS$_2$\cite{Qian2014}  & WTe$_2$ & $\nu=1$ & 51    \\ 
      MoSe$_2$\cite{Qian2014} & WTe$_2$ & $\nu=1$ & 41    \\ 
      WSe$_2$\cite{Qian2014}  & WTe$_2$ & $\nu=1$ & 32   \\ 
      OsCl$_3$\cite{Sheng2017} & BiI$_3$ & $C=1$  & 64  \\ 
      GeTe$_2$\cite{Zhang2018a} & MnS$_2$   & $\nu=1$ & 32 \\ 
      SnS\cite{Liu2015c}   & PbS       & $C_M=2$ & 67 \\ 
      SnSe\cite{Liu2015c}     & PbS       & $C_M=2$ & 84 \\ 
      SnTe\cite{Liu2015c}     & PbS       & $C_M=2$ & 28 \\ 
      PbS\cite{Liu2015c}      & PbS       & $C_M=2$ & 422 \\ 
      PbSe\cite{Liu2015c}     & PbS       & $C_M=2$ & 478 \\ 
      PbTe\cite{Liu2015c}     & PbS       & $C_M=2$ & 271 \\ 
    \end{tabular}
    \caption{Overview of known topological materials found by computational screening. The topology is specified by either the Chern number $C$, the $\mathbb{Z}_2$ index $\nu$ or the mirror Chern number $C_M$. We also state the calculated Kohn-Sham gap (KS gap).}
  \end{center}
\end{table}
Tab. I contains known topological 2D materials.  We find most of the materials that have previously been predicted to exhibit a non-trivial band topology. For example, graphene\cite{Kane2005a} and its derivatives silicene, germanene, and stanene\cite{Ezawa2015} as well as the transition metal dichalcogenides MoS$_2$, MoSe$_2$, and WSe$_2$ in the 1T' phase (WTe$_2$ crystal structure).\cite{Qian2014} However, some well-known 2D topological insulators are missing from this table. For example, WTe$_2$ and WS$_2$ in the 1T' phase. These materials are present in C2DB, but are semi-metals in the PBE approximation\cite{Qian2014} and are therefore excluded from the present compilation. In addition CoBr$_2$ in the CdI$_2$ prototype has previously been predicted to be a QAHI based on PBEsol,\cite{Chen2017a} which we have confirmed but the material is metallic within PBE and is therefore not included here. On the other hand, we find both HgSe and HgTe to be QSHIs although these materials have previously been reported to be trivial insulators based on calculations with a modified Becke-Johnson LDA functional.\cite{Li2015}

\begin{table*}[tb]
  \begin{center}
    \label{tab:table2}
    \begin{tabular}{l|c|c|c|c|c|c}
      Material & Prototype & Topology & KS gap [meV] & HOF  [eV]  & EACH [eV]\\ 
      \hline
      AuCl     & FeSe    & $\nu=1$ & 20    & 0.10 & 0.30 \\
      \bf{CrAsBi}   & BiTeI   & $\nu=1$ & 35  & 0.31 & 0.46 \\
      IrSe     & GaSe    & $\nu=1$ & 134 & 0.18 & 0.45 \\
      \bf{TiTe}     & GaSe    & $\nu=1$ & 109 & -0.08 & 0.63 \\
      \bf{ZrTe}     & GaSe    & $\nu=1$ & 207 & -0.28 & 0.65 \\
      AuI$_3$  & BiI$_3$ & $\nu=1$ & 109  & 0.10 & 0.10 \\
      \bf{TiIN}     & FeOCl & $\nu=1$ & 62  & -1.18 & -0.26 \\
      TlClSe   & FeOCl & $\nu=1$ & 27 & -0.36 & 0.32 \\
      \bf{TiS}      & CuI       & $\nu=1$ & 54 & -1.13 & 0.31 \\ 
      \bf{TiCl}     & CuI       & $\nu=1$ & 13 & -0.64 & 0.45 \\ 
      \bf{ZrS}      & CuI       & $\nu=1$ & 132 & -1.16 & 0.26 \\ 
      \bf{ZrSe}      & CuI       & $\nu=1$ & 20 & -0.91 & 0.25 \\ 
      \bf{ZrCl}      & CuI       & $\nu=1$ & 37 & -0.59 & 0.73 \\  
      \bf{ZrBr}      & CuI       & $\nu=1$ & 45 & -0.34 & 0.68 \\ 
      \bf{SbCl}      & CuI       & $\nu=1$ & 434 & -0.46 & 0.13 \\  
      \bf{SbBr}      & CuI       & $\nu=1$ & 442 & -0.31 & 0.10 \\ 
      \bf{SbI}       & CuI       & $\nu=1$ & 584 & -0.12 & 0.13 \\ 
      \bf{HfS}       & CuI       & $\nu=1$ & 158 & -0.89 & 0.43 \\ 
      \bf{HfSe}      & CuI       & $\nu=1$ & 42 & -0.64 & 0.54 \\ 
      \bf{ReS}       & CuI       & $\nu=1$ & 309 & 0.10 & 0.54 \\ 
      \bf{HgCl}      & CuI       & $\nu=1$ & 129 & -0.37 & 0.23 \\ 
      \bf{HgBr}      & CuI       & $\nu=1$ & 188 & -0.25 & 0.22 \\ 
      \bf{PbF}       & CuI       & $\nu=1$ & 116 & -1.43 & 0.45 
    \end{tabular}
    \caption{Overview of new QSHIs without mirror symmetry. All the materials have a $\mathbb{Z}_2$ index of $\nu=1$. We also state the calculated Kohn-Sham gap (KS gap), the heat of formation (HOF) and the energy above the convex hull (EACH). The dynamically stable materials are shown in bold face.}
  \end{center}
\end{table*}

\begin{table*}[tb]
  \begin{center}
    \label{tab:table3}
    \begin{tabular}{l|c|c|c|c|c|c}
      Material & Prototype & Topology & KS gap [meV] & HOF  [eV]  & EACH [eV]\\ 
      \hline
      HgO      & BN      & $\nu=1$, $C_M=1$ & 302 & -0.15 & 0.21 \\
      \bf{PdSe$_2$} & MoS$_2$ & $\nu=1, C_M=1$ & 229  & -0.02 & 0.27 \\
      \bf{AuTe}     & GaS     & $\nu=1, C_M=1$ & 37 & 0.03 & 0.11 \\
      WO       & GaS     & $\nu=1, C_M=1$ & 53 & -1.01 & 0.49 \\
      \bf{RhO}      & GaS     & $\nu=1, C_M=3$ & 67 & -0.38 & 0.25 \\
      \bf{IrO}      & GaS     & $\nu=1, C_M=3$ & 122 & -0.16 & 0.51 \\
      ReI$_3$  & AgBr$_3$ & $\nu=1, C_M=1$  & 141 & 0.37 & 0.43 \\
      ReCl$_3$ & TiCl3$_3$ & $\nu=1, C_M=1$ & 220 & -0.18 & 0.52 \\
      \bf{WI$_3$}   & TiCl3$_3$ & $\nu=1, C_M=1$ & 222 & 0.19 & 0.19 \\
      \bf{GeTe$_2$} & MnS$_2$   & $\nu=1, C_M=1$ & 32 & 0.27 & 0.33 \\
      \bf{RuTe$_2$} & MnS$_2$   & $\nu=1, C_M=1$ & 157 & 0.34 & 0.68 \\
      \bf{OsS$_2$}  & MnS$_2$   & $\nu=0, C_M=2$ & 111 & 0.25 & 0.63\\
      OsSe$_2$ & MnS$_2$   & $\nu=0, C_M=2$ & 144 & 0.51 & 0.68\\
      OsTe$_2$ & MnS$_2$   & $\nu=0, C_M=2$ & 117 & 0.70 & 0.77\\
    \end{tabular}
    \caption{Overview of new crystalline topological insulators protected by mirror symmetry. The topology is specified the mirror Chern number $C_M$. All the materials are invariant under time-reversal symmetry and the associated $\mathbb{Z}_2$ index thus becomes $\nu=1$ if $C_M$ is odd and $\nu=0$ if $C_M$ is even. We also state the calculated Kohn-Sham gap (KS gap), the heat of formation (HOF) and the energy above the convex hull (EACH). The dynamically stable materials are shown in bold face.}
  \end{center}
\end{table*}

\begin{table*}[tb]
  \begin{center}
    \begin{tabular}{l|c|c|c|c|c|c}
      Material & Prototype & Topology & KS gap [meV] & HOF  [eV]  & EACH [eV]\\ 
      \hline
      \bf{PdF$_2$}  & GeS$_2$ & $C=-2$ & 27  & -0.83 & 0.38 \\
      \bf{FeCl$_3$} & BiI$_3$ & $C=1$  & 13  & -0.66 & -0.08\\
      \bf{FeBr$_3$} & BiI$_3$ & $C=1$ & 42  & -0.38 & -0.04 \\
      \bf{PdI$_3$}  & BiI$_3$ & $C=-1$ & 36  & -0.13 & 0.06\\
      \bf{CoBr$_3$} & AgBr$_3$ & $C=-2$  & 27  & -0.21 & -0.16\\
      \bf{MoS$_2$}  & MnS$_2$   & $C=2$ & 133 & 0.10 & 1.05 \\
    \end{tabular}
    \caption{Overview of new QAHIs found by computational screening. The topology is specified by either the Chern number $C$. We also state the calculated Kohn-Sham gap (KS gap), the heat of formation (HOF) and the energy above the convex hull (EACH). The dynamically stable materials are shown in bold face.}
  \end{center}
\label{tab:table4}
\end{table*}

Tabs. II-IV contains all the topological insulators that to our knowledge have not been reported prior to this work. The dynamically stable compounds (shown in bold face) are likely to be the only ones that are experimentally relevant, but we include all materials that we have found for completeness. 

\begin{figure}[b]
   \includegraphics[width=7.5 cm]{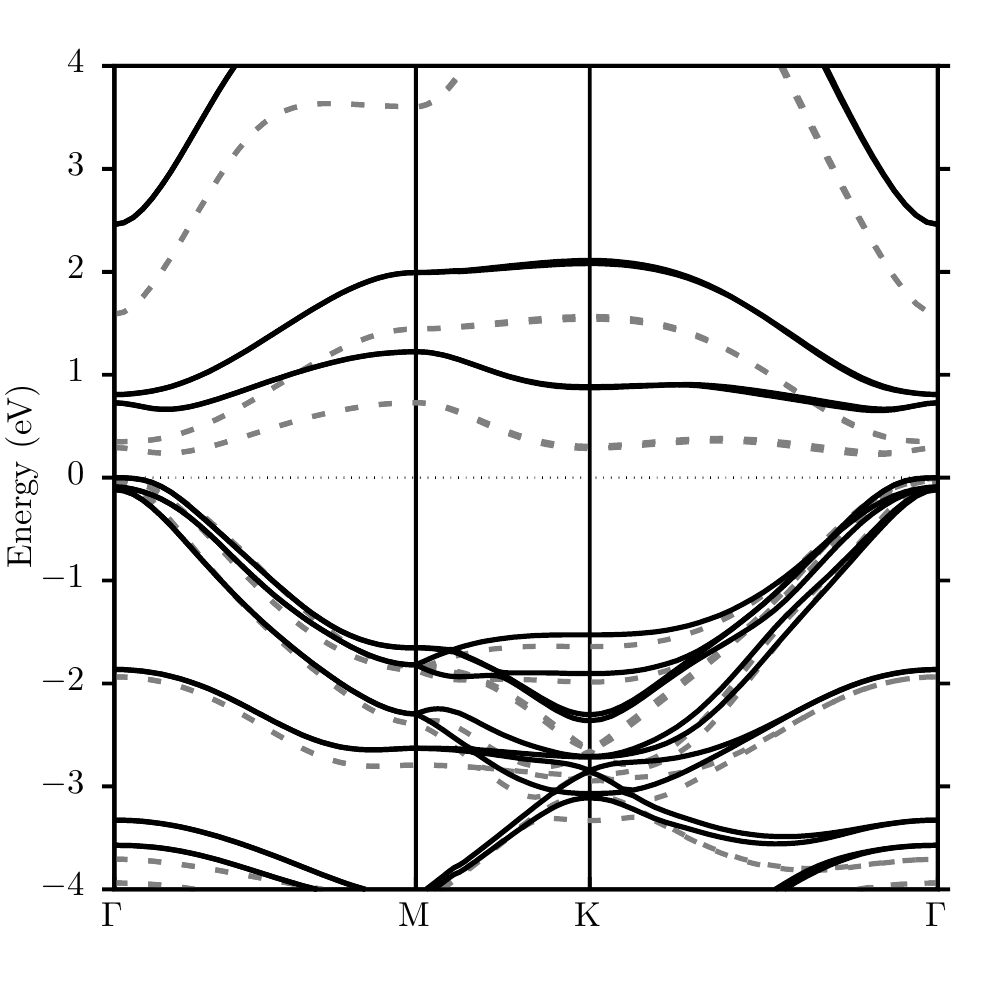}
\caption{LDA (dashed) and GW (solid) band structures of PdSe$_2$ in the MoS2 prototype. In both cases the energy of the top of the valence band has been set to zero. The metallic point, where the material undergoes a topological phase transition is indicated by a vertical line.}
\label{fig:PdSe2}
\end{figure}

We find 27 new $\mathbb{Z}_2$ topological insulators that are dynamically stable (Tabs. II and III). We start by noting that the list is dominated by a 15 materials in the CuI crystal structure, with the antimony halides exhibiting band gaps exceeding 0.4 eV. Among the remaining materials, PdSe$_2$ in the MoS$_2$ crystal structure is the one with the largest band gap of 0.23 eV. However, it is situated 0.25 eV/atom above the convex hull with two other materials of the same stoichiometry being more stable. Nevertheless, the difference in stability is similar to the difference between MoS$_2$ in the 2H phase and the 1T' phase, which are both accessible by modern synthesis techniques. To test the reliability of the topological gap obtained with PBE, we have calculated LDA and $GW$ band structures fully including spin-orbit coupling in the self-energy. The result is shown in Fig. \ref{fig:PdSe2}. The LDA gap is 0.225 eV, which is very similar to the PBE value. However, the $GW$ gap is 0.65 eV, which nearly comprises a threefold increase of the Kohn-Sham gap. In fact, to our knowledge this is one of the largest gaps reported for a two-dimensional topological insulator. A similar dramatic increase was recently reported for a Jacutingaite crystal structure where PBE yielded a gap of 0.15 eV and GW predicted a gap of 0.5 eV.\cite{Marrazzo2017a} Such a dramatic increase in band gap has not been reported for three-dimensional topological insulators and could be related to the reduced screening in 2D. In principle, the predicted topology of the materials may depend on the approximation used to obtain the eigenstates. For example, in Ref. \onlinecite{Vidal2011} it was that shown that DFT can lead to a False-positive conclusion for the non-trivial topology of certain 3D materials and GW calculatians may reverse the band inversion leading to the predicted non-trivial topology in DFT. However, we will not perform full GW calculations (with spin-orbit coupling) for all the materials in the present work, but note that false-positives could be a caveat for the present method. We discuss this issue further below in the context of QAHIs.

In addition to the 6 well-known mirror crystalline topological insulators displayed in Tab. \ref{tab:table1}, we also find 3 osmium dichalcogenides in the MnS$_2$ crystal structure with a mirror Chern number of $C_M=2$, which are displayed in Tab. III. Only one of these - OsO$_2$ - is stable though. As noted previously, any $\mathbb{Z}_2$ odd topological insulator with mirror symmetry must have an odd Chern number. In those cases the $\mathbb{Z}_2$ index does not exhaust the topological properties and we may distinguish the topological classes corresponding to $C_M=1$ and $C_M=3$ for example as in the case of RhO and AuTe in the GaS crystal structure. Again the physical consequences only emerge when considering an edge where the difference in Chern numbers would yield the number of protected gapless edge states. In general, one would not expect an interface between two $\mathbb{Z}_2$ topological insulators ($\nu=1$) to exhibit gapless interface modes. However, any interface between RhO and AuTe any interface that conserves the mirror symmetry would host four topologically protected gapless edge modes in - two for each mirror sector.

We would also like to emphasize that the screening has resulted in 6 new candidates for quantum anomalous Hall insulators, which are displayed in Tab. IV. This is of particular interest, since an experimental demonstration of the quantum anomalous Hall effect in a pristine 2D material is still lacking. Specifically, the materials FeCl$_3$ and FeBr$_3$ are highly stable and situated less than 0.1 eV/atom above the convex hull. However, the band structure of FeBr$_3$ shown in Fig. \ref{fig:FeBr3} exhibits rather flat bands and indicates that the electrons in these materials are strongly correlated. It is thus likely that the PBE band gap provides a poor estimate of the fundamental gap of the material and even the topological properties could be wrong if PBE does not describe the band inversion correctly. In order to test the reliability of PBE we have tested the topological properties with various semi-local functionals and with PBE+U. Using LDA, RPBE, and PBEsol yield a trivial topology whereas PBE and revPBE predicts a QAHI with $C=1$. Using PBE+U we obtain a QAHI for values of U below 0.18 eV and a trivial insulator for U$>0.18$ eV. In all cases the geometry was optimized with the given functional. The topological properties of FeBr$_3$ are thus highly sensitive to the method used and with the methodology applied here it is not possible to determine whether or not the material is a QAHI.

\subsection{Magnetic anisotropy}
The quantum anomalous Hall effect is driven by spin-orbit mediated band inversion in magnetic materials. Moreover, as a consequence of the Mermin-Wagner theorem magnetic order cannot exist without magnetic anisotropy in 2D materials and spin-orbit effects thus have another crucial role to play for these materials. In fact, the Curie temperature in a 2D ferromagnet is strongly dependent on the anisotropy as well as the exchange coupling constants and the first example of 2D ferromagnetic order was only observed very recently in the trivial insulator CrI$_3$. Some of the present authors have shown that the Curie temperatures in 2D can be obtained from Monte Carlo simulations\cite{Torelli2018} based on the classical Heisenberg model
\begin{equation}
 H=-\frac{1}{2}J\sum_{\langle ij\rangle}\mathbf{S}_i\cdot\mathbf{S}_i-A\sum_i(S^z_j)^2-\frac{1}{2}B\sum_{\langle ij\rangle}S^z_iS^z_j,
\end{equation}
where $\langle ij\rangle$ denotes sum over nearest neighbors. The exchange, and anisotropy constants $J$, $A$, and $B$ can be obtained from first principles calculations combined with an energy mapping scheme that assumes magnetic moments to be located on transition metal atoms.\cite{Olsen2017, Torelli2018} Since the magnetic structure is typically isotropic in the plane of the material a minimal requirement for magnetic order at finite temperatures is an out-of-plane easy axis. For the magnetic materials in Tab. IV we find $T_c=2 K$ and $T_c=274K$ respectively for FeBr$_3$, CoBr$_3$. The remaining magnetic materials are predicted to lack magnetic order at any finite temperature due to an in-plane easy axis or, more precisely, a negative spin-wave gap.\cite{Torelli2018} From these calculations CoBr$_6$ appears highly promising. However, additional calculations shows that this material is more stable in the BiI$_3$ crystal structure (similar to CrI$_3$), which is non-magnetic and a trivial insulator. We note that the materials with an in-plane easy axis that are excluded here could in general give rise to a finite critical temperature if there is additional small in-plane anisotropy such that the rotational symmetry is explicitly broken. It is, however, rather difficult to predict the critical temperature in those cases and we expect critical temperatures to be low due to the approximate in-plane anisotropy. One exception to this may be provided by OsCl$_3$\cite{Sheng2017}, which has a rather large magnetic anisotropy of $\sim40$ meV per Os atom.

\begin{figure}[b]
   \includegraphics[width=7.5 cm]{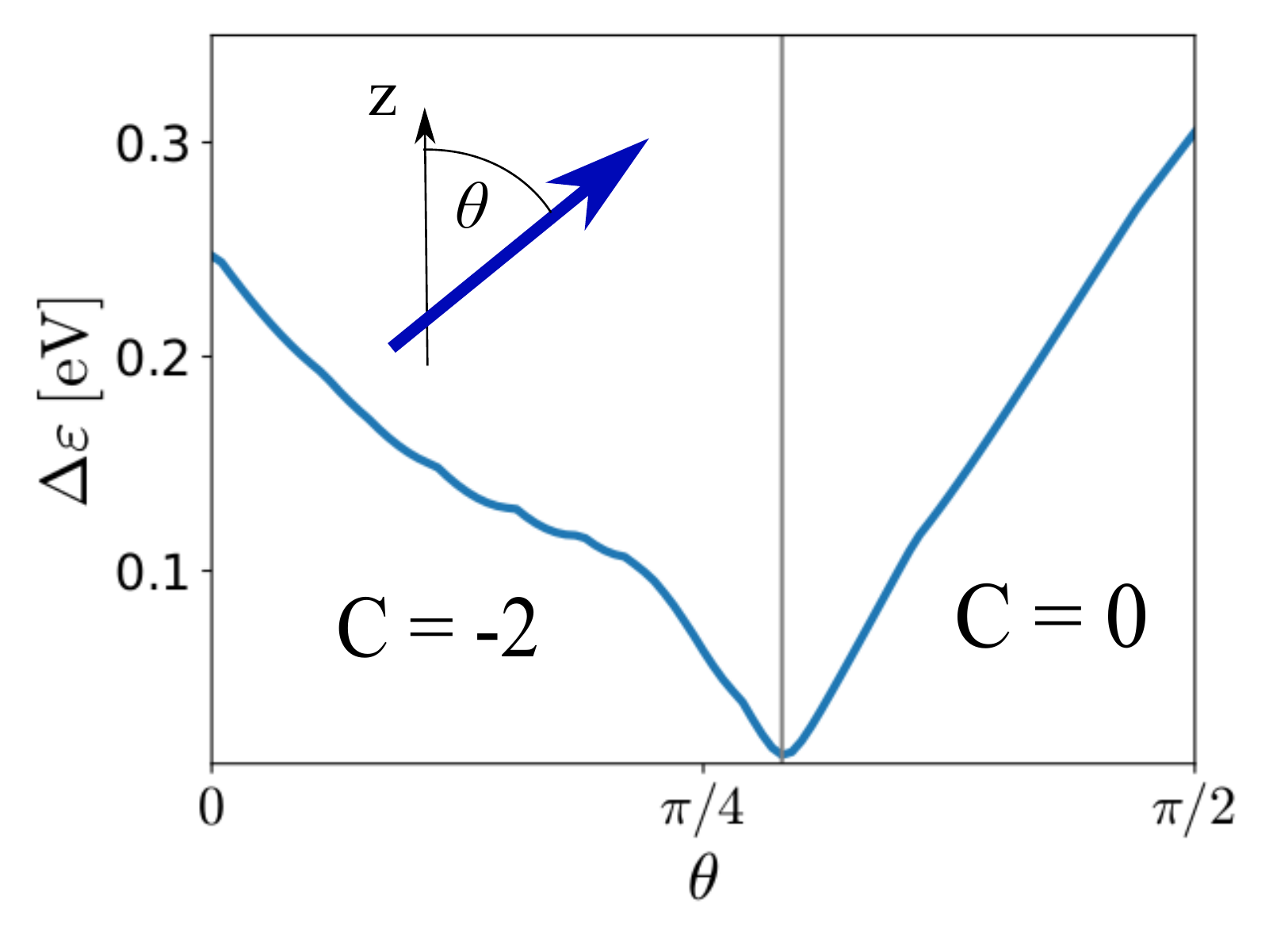}
\caption{Band gap as a function of angle with the out-of-plane axis in OsO$_2$. The material undergoes a topological transition for $C=0$ to $C=-2$ when the gap closes in the vicinity of $\pi/4$.}
\label{fig:OsO2}
\end{figure}

It is interesting to note that small gap magnetic insulators with large spin-orbit coupling may undergo topological phase transitions upon a rotation of the spin structure. As an example we take OsO$_2$ in the CdI$_2$ crystal structure, which has an in-plane easy axis and is not found in Tab. IV because it is a trivial insulator. However, if the magnetic moments are rotated out of plane to align with the axis perpendicular to the plane of the material it becomes a QAHI with a Chern number of $C=-2$. This is only possible if the band gap closes at some point when the magnetic moments are rotated from the in-plane to the out-of-plane configuration. In Fig. \ref{fig:OsO2} we show the band gap as a function of the polar angle $\theta$ that the magnetic moments form with the $z$-axis. The band gap is seen to close in the vicinity of $\theta=\pi/4$ where the system undergoes a topological phase transition from $C=0$ to $C=2$. Such a rotation can be accomplished by applying a magnetic field and comprises a mechanism under which the gapless edge states can switched on or off by external means.

\section{Discussion}\label{sec:discussion}
We have implemented and performed an automated search for topologically non-trivial 2D materials in the Computational 2D Database. The method is based on a direct evaluation of Berry phases from the Kohn-Sham states and circumvents the common mapping to tight binding models via Wannier functions\cite{Gresch2017} that can sometimes make automation cumbersome. 

In addition to several well-known topological insulators we have found 45 new materials of which, 18 are predicted to be stable. Of particular interest are the 6 magnetic QAHI. The experimental demonstration of the quantum anomalous Hall effect in a pristine 2D material would constitute a major breakthrough in the field of topological materials science. However, even if any of these materials could be synthesized the experimental verification of the effect will be highly tricky, since all of the experimentally relevant QAHIs have band gaps below 0.1 eV. Moreover, we have shown that the theoretical prediction of the topological properties is non-trivial due to strong correlation and the predictions based on PBE calculations may not be reliable as we exemplified in the case of FeBr$_3$. Nevertheless, the PBE predictions for QAHIs presented here provide indications that some of these materials could be highly interesting to put under experimental scrutiny. Most notable CoBr$_3$ in the AgBr$_3$ crystal structure, which we predict to have a Curie temperature of 274 K, but which is also predicted to be less stable compared to the similar BiI$_3$ crystal structure.

The 10 stable QSHIs found have PBE band gaps between 0.05 and 0.23 eV. We have only performed full spinorial GW calculations for PdSe$_2$ where we found a threefold increase of the band gap. It would be highly interesting to perform carefully converged spinorial GW calculations for all the topological insulators presented in Tabs. I-IV, but this is beyond the scope of the present work. However, based on the cases of PdSe$_2$ and jacutegaite\cite{Marrazzo2017a} we believe it is likely that several of the materials could have significantly larger gabs than predicted by PBE. None of the materials in Tabs. II-IV have been synthesized yet. But considering the rapid pace at which experimental techniques are currently evolving, we expect an experimental realization of one or several of the predicted topological insulators should be within reach in near future.

\section{Acknowledgments}
This project has received funding from the European Research Council (ERC) under the European Union's Horizon 2020 research and innovation programme (grant agreement No 773122, LIMA). T. Olsen and D. Torelli was supported by the Independent Research Fund Denmark, grant number 6108-00464A. T. Okugawa acknowledges support from the Scandinavia-Japan Sasakawa Foundation. T. Deilmann acknowledges financial support from the Villum foundation.


%

\end{document}